\documentclass[final,2p,times,twocolumn]{elsarticle}

\usepackage{amssymb}

\begin{document}

\begin{frontmatter}



\title{Self-similarity in jet events following from p-p collisions at LHC}


\author[NCBJ]{Grzegorz Wilk}
\ead{wilk@fuw.edu.pl}


\address[NCBJ]{National Centre for Nuclear Research,
        Department of Fundamental Research, Ho\.za 69, 00-681
        Warsaw, Poland}

\author[JKU]{Zbigniew W\l odarczyk}
\ead{zbigniew.wlodarczyk@ujk.edu.pl}

\address[JKU]{Institute of Physics, Jan Kochanowski University,
\'Swi\c{e}tokrzyska 15, 25-406 Kielce, Poland}

\begin{abstract}
Using a Tsallis nonextensive approach, we simultaneously analyze
recent data obtained by the LHC ATLAS experiment on distributions
of transverse momenta of jets, $p_T^{jet}$, together with
distributions of transverse momenta of particles produced within
these jets (defined relative to the jet's axis), $p_T^{rel}$, and
their multiplicity distributions, $P\left( N\right)$. The
respective nonextensivity parameters for distributions of jets,
$q_{jet}$, for distributions of particles in jets, $q_{rel}$ and
the global nonextensivity parameter obtained from $P\left( N
\right)$, $q_N$, were then compared with nonextensivity parameters
$q$ obtained from minimum bias $pp$ collisions at energies
corresponding to the energies of these jets. The values of the
corresponding nonextensivity parameters were found to be similar,
strongly indicating the existence of a common mechanism behind all
these processes. We tentatively identify this as a {\it
self-similarity} property known to be present there and resulting
in Tsallis type distributions. If confirmed, this would
considerably strengthen the nonextensive Tsallis approach.
\end{abstract}


\end{frontmatter}

\vspace{1cm}

\section{Introduction}\label{sec:I}

For some time now it is known that transverse momentum spectra of
different kinds measured in multiparticle production processes,
which change character from exponential at small values of $p_T$
to power-like at large $p_T$, can be described by a simple
two-parameter formula,
\begin{equation}
h\left( p_T\right) = C \left( 1 + \frac{p_T}{nT}\right)^{-n}.
\label{eq:FitTn}
\end{equation}
This was first proposed in \cite{Oldpar} as the simplest formula
extrapolating the large $p_T$ power behavior expected from parton
collisions to exponential behavior observed for $p_T \rightarrow
0$. At present it is known as the QCD-based {\it Hagedorn formula}
\cite{H} and was used in many fits to recent data. However, in
many branches of physics Eq. (\ref{eq:FitTn}), with $n$ replaced
by $n = 1/(1-q)$, is more widely known as the Tsallis formula
\cite{Tsallis}. In this case, $q$ is known as a nonextensivity
parameter. In this form, Eq. (\ref{eq:FitTn}) is usually supposed
to represent a nonextensive generalization of the Boltzmann-Gibbs
exponential distribution, $\exp( - p_T/T)$, used in a statistical
description of multiparticle production processes, with $q$ being
a new parameter, in addition to previous "temperature" $T$. Such
an approach is known as nonextensive statistics \cite{Tsallis} in
which the parameter $q$ summarily describes all features causing a
departure from the usual Boltzmann-Gibbs statistics (in particular
it can be shown that it is directly related to the possible
intrinsic, nonstatistical fluctuations of the temperature $T$
\cite{qWW,WW}). However, the Tsallis distribution also emerges
from a number of other more dynamical mechanisms, for example see
\cite{cond} for more details and references. In all possible
scenarios leading to Eq. (\ref{eq:FitTn}), the "temperature", or,
in general, scale parameter $T$, is given by the mean value of the
transverse momentum, $\langle p_T \rangle = 2nT/(n - 3)$, and we
do not discuss here its possible dependence on energy and the
nonextensivity parameter. For large values of transverse momenta,
$p_T >> nT$, Eq. (\ref{eq:FitTn}) becomes scale free (independent
of $T$) distribution. The Tsallis distribution was successfully
used for a description of all kinds of multiparticle production
processes in a wide range of incident energy (from few GeV up to
few TeV) and in a broad range of transverse momenta (see, for
example, reviews \cite{cond,WW}. In particular, it turned out that
it also successfully describes transverse momenta of charged
particles measured by LHC experiments, the flux of which changes
by over $15$ orders of magnitude \cite{CYWGW}\footnote{In
\cite{CYWGW1} these results were derived from QCD considerations.
It turns out that, although one gets a Tsallis-like formula, there
is a $p_T$ dependent prefactor, the presence of which affects the
value of the $q$ parameter. Also, in the low $p_T$ domain, Tsallis
distribution with $p_T^2$ seems to do better than the one with
$p_T$. Both choices are possible, depending on the circumstances,
cf. \cite{Rodos} for details. In our case both would result in the
same conclusions.}.

The Tsallis distribution was recently used in an analysis of the
distribution of the longitudinal component of momenta of particles
within jets produced in $pp$ collisions \cite{Biro1} which, from
this point of view, is similar to what was found in $e^+e^-$
collisions \cite{Biro2}. Recent ATLAS data \cite{ATLAS,ATLAS1}
allow us to extend such an analysis to transverse characteristics
of jets and charged particles within them. This is because they
provide both the distributions of transverse momenta of jets
produced at LHC energies, $p_T^{jet}$, and distributions of
transverse momenta of particles produced within these jets
(defined relative to the jets), $p_T^{rel}$. One can then retrieve
and discuss the respective nonextensivity parameters of jets,
$q_{jet}$, and particles produced within them, $q_{rel}$. In
addition, because \cite{ATLAS} at the same time also provides
multiplicity distributions within jets, $P(N)$, it is possible to
confront both nonextensivities with that obtained from an analysis
of $P(N)$, $q_N$. This is the subject of the present
work\footnote{The other two LHC experiments, ALICE and CMS, do not
provide such results for the same experimental conditions and
using the same criteria for data selection.}.

\section{Transverse momentum distributions of jets and particles
within jets} \label{sec:pT}

In what follows we shall concentrate on ATLAS data \cite{ATLAS}.
They were taken at energy $7$ TeV and in rapidity window $|y| <
1.9$ measured jets observed in very narrow jet cones defined by
$R=\sqrt{\Delta \eta^2 + \Delta \phi^2}$ (where $\Delta \phi$ and
$\Delta \eta$ are, respectively, the azimuthal angle and the
pseudorapidity of the hadrons relative to that of the jet, $\eta =
- \ln \tan \theta$, with $\theta$ being the polar angle), namely
$R = 0.6$. Distributions of transverse momenta, $p_T^{jest}$, of
jets of charged particles were observed,
\begin{equation}
f\left(p^{jet}_T\right) =
\frac{1}{N_{jet}}\frac{dN_{jet}}{dp^{jet}_T} \label{eq:jetdistr}
\end{equation}
and also distributions of transverse momenta
\begin{equation}
p^{rel}_T = \frac{\left | \vec{p}\times \vec{p_{jet}}\right |}
{\left | \vec{ p_{jet}}\right |} \label{eq:ptpart}
\end{equation}
of all $N$ particles (only charged) in the jet,
\begin{equation}
f\left(p^{rel}_T\right) = \frac{1}{N}\frac{dN}{dp^{rel}_T}.
\label{eq:ptpartdistr}
\end{equation}
In addition, \cite{ATLAS} also provides multiplicity distributions
of particles produced within observed jets, $P(N)$.

\begin{figure}[h]
\includegraphics[width=8cm]{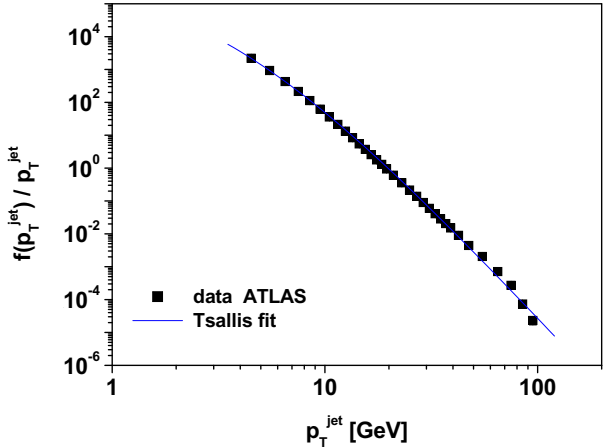}
\vspace{-10mm} \caption{(Color online) Distribution of $p_T^{jet}$
for jets at $\sqrt{s} = 7$ TeV fitted by Tsallis distribution
(\ref{eq:FitTn}) with $T = 0.45$ GeV and $n = 7$ (which
corresponds to $q = 1.14$). Data are taken from \cite{ATLAS}.}
\label{Fig1}
\end{figure}

It should be stressed that the pure power law distribution,
$f\left( p_T\right) \sim p_T^{-\gamma}$, is not experimentally
observed for jets. The observed slope parameter $\gamma$ depends
on $p_T$, $\gamma = \gamma\left( p_T\right)$. However, a Tsallis
distribution (\ref{eq:FitTn}) emerges if one accounts for this
dependence and assumes it in the following two parameter ($n$ and
$T$) form,
\begin{equation}
\gamma\left( p_T \right) =
          n \frac{\ln\left( nT + p_T\right)}{\ln \left(
          p_T\right)}  +  \frac{ (n - 1)\ln(nT) + \ln(n-1)}{ \ln \left(
          p_T\right)}. \label{eq:qnT}
\end{equation}
In this case, the transverse momentum distribution for jets can be
fitted by a Tsallis formula (\ref{eq:FitTn}) with $n \simeq 7$ and
$T = 0.45$ GeV, cf. Fig. \ref{Fig1}.

\begin{figure}[h]
\includegraphics[width=8cm]{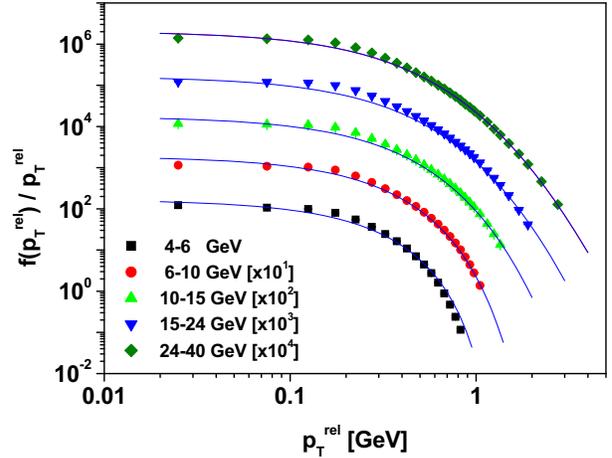}
\vspace{-10mm} \caption{(Color online) Distributions of
$p^{rel}_T$ particles inside the jets with different values of
$p_T^{jet}$ obtained in \cite{ATLAS}, fitted using Tsallis
distribution (\ref{eq:FitTn}). To make distributions readable, the
consecutive curves $i = 0,1,2,\dots$ were multiplied by $10^i$.
For all curves $T = 0.18$ GeV. The corresponding values of the
parameter $n$ (and $q = 1+1/n$) are listed in Table\ref{Table1}.
Data are taken from \cite{ATLAS}.} \label{Fig2}
\end{figure}
\begin{figure}[h]
\includegraphics[width=8.5cm]{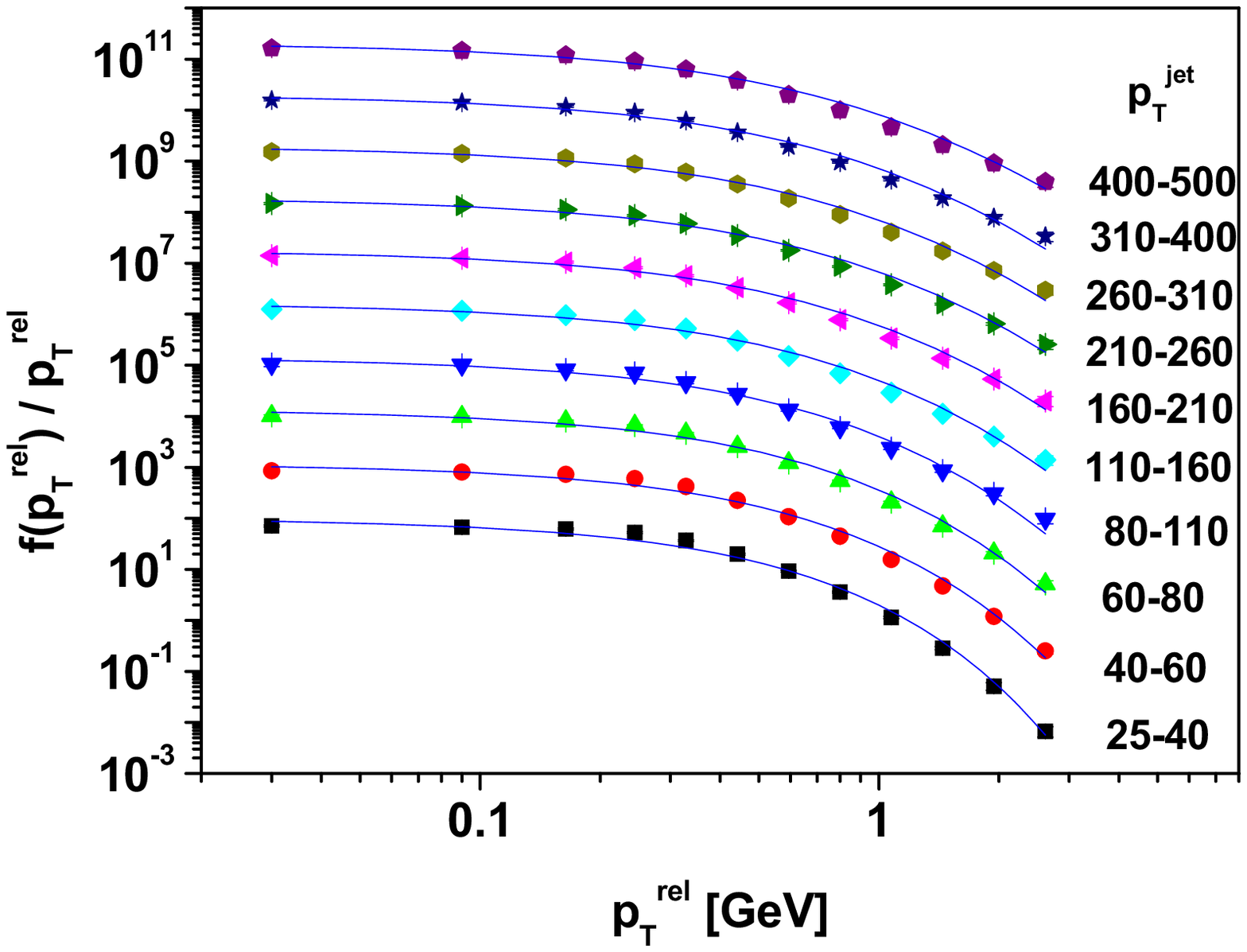}
\vspace{-10mm} \caption{(Color online) The same as in Fig.
\ref{Fig2} but now for jests with larger $p_T^{jet}$ for which all
curves have $T = 0.25$ GeV. The corresponding values of the
parameter $n$ (and $q = 1+1/n$) are given in Table \ref{Table2}.
Data are taken from \cite{ATLAS1}.} \label{Fig3}
\end{figure}

Data on distributions of transverse momenta $p_T^{rel}$ of
particles produced within the jet are presented in two papers. In
\cite{ATLAS} are data for $p^{jet}_T \leq 40$ GeV and in
\cite{ATLAS1} for $p_T^{jet} > 40$ GeV. All can be fitted by a
Tsallis formula (\ref{eq:FitTn}) and results are shown in Fig.
\ref{Fig2} and Table \ref{Table1}, for the first set, and in Fig.
\ref{Fig3} and Table \ref{Table2}, for the second one. It must be
stressed at this point that the uncertainty is large in getting
the precise values of parameters $T$ and $n$ from fits because
both variables are correlated. One also has to remember that data
from \cite{ATLAS1} presented in Fig. \ref{Fig3} differ from those
from \cite{ATLAS} and presented in Fig. \ref{Fig2}. Namely, they
were collected for $|\eta| < 1.2$ and $p_T^{track} > 0.5$ GeV (to
be compared with $|\eta| < 1.9$ and $p_T^{track} > 0.3$ GeV in the
former case). This fact influences multiplicity in jets (which is
smaller in the later case), which, in turn, influences the value
of the parameter $T$.

\begin{table}[h]
\caption[Table1]{Fit parameters for Fig. \ref{Fig2}; $q = 1 +
1/n$.}
\begin{center}
\begin{tabular}{|r|r|r|r|}
\hline $p_T^{jet}$ [GeV] & $T$ [GeV] & $n$~~~&$q$~~~~~ \\
\hline

 4 -  6~~~ & 0.18~~~ & -8.5~ & 0.88~~~ \\
 6 - 10~~~ & 0.18~~~ & -17~~~& 0.94~~~ \\
10 - 15~~~ & 0.18~~~ &  55~~ & 1.02~~~ \\
15 - 24~~~ & 0.18~~~ &  16~~ & 1.06~~~ \\
24 - 40~~~ & 0.18~~~ & 11.5~~& 1.09~~~ \\
\hline
\end{tabular}
\label{Table1}
\end{center}
\end{table}

\begin{table}[h]
\caption[Table2]{Fit parameters for Fig. \ref{Fig3}; $q = 1 +
1/n$.}
\begin{center}
\begin{tabular}{|r|r|r|r|}
\hline $p_T^{jet}$ [GeV] & $T$ [GeV] & $n$~~~&$q$~~~~ \\
\hline

 25 - 40~~~ & 0.25~~~ & 70~~ & 1.014~~ \\
 40 - 60~~~ & 0.25~~~ & 25~~ & 1.040~~ \\
 60 - 80~~~ & 0.25~~~ & 18~~ & 1.056~~ \\
 80 - 110~~~ & 0.25~~~ & 15~~ & 1.067~~ \\
110 - 160~~~ & 0.25~~~ & 12~~ & 1.083~~ \\
160 - 210~~~ & 0.25~~~ & 10~~ &  1.100~~   \\
210 - 260~~~ & 0.25~~~ &  9~~~ & 1.111~~   \\
260 - 310~~~ & 0.25~~~ &   9~~~ & 1.111~~  \\
310 - 400~~~ & 0.25~~~ &   9~~~ & 1.111~~  \\
400 - 500~~~ & 0.25~~~ &   7.5~ & 1.133~~   \\
\hline
\end{tabular}
\label{Table2}
\end{center}
\end{table}

Notice the negative values of the parameter $n$ (or,
correspondingly, $q < 1$ values of the nonextensivity parameter)
for small values of the $p_T^{jet}$, i.e., for small values of the
energy of such jets seen in Fig. \ref{Fig2}. This fact is
connected with the limitation of the available phase space in this
case. Actually, maximal values for the ratios
$p_T^{rel}/p_T^{jet}$ for data in Fig. \ref{Fig2} are in the range
$0.09~-~0.15$ and in Fig. \ref{Fig3} in the range $0.006~-~0.09$.
The nonextensivity parameter drops below unity for distributions
with $p_T^{rel}/p_T^{jet} > 0.12$.

\section{Multiplicity distributions within jets}
\label{sec:N}

From our experience with applications of Tsallis statistics to
multiparticle production processes, we know \cite{Nq} that
multiplicity distribution of particles energy spectra of which
follow Tsallis distribution has Negative Binomial form
(NBD)\footnote{Cf., also \cite{DdDetal} where similar results were
obtained from apparently different point of views. In fact there
is a parameter equivalent to $q$ and a resulting distribution can
be written in Tsallis form.},
\begin{equation}
P(N) =
\frac{\Gamma(N+k)}{\Gamma(N+1)\Gamma(k)}\frac{\left(\frac{\langle
N\rangle}{k}\right)^N}{\left(1 + \frac{\langle N\rangle}{k}
\right)^{k + N}}, \label{eq:NBD}
\end{equation}
with
\begin{equation}
\frac{1}{k} = \frac{Var(N)}{\langle N\rangle^2} - \frac{1}{\langle
N\rangle} = q_N - 1. \label{eq:kq}
\end{equation}
Whereas for NBD $q > 1$ and parameter $k$  in (\ref{eq:kq}) is
positive, for the $q < 1$ case $k$ becomes negative ($k
\rightarrow - \kappa$) and NBD becomes a binomial distribution
(BD),
\begin{equation}
P(N) = \frac{\Gamma(\kappa + 1)}{\Gamma(N+1)\Gamma(\kappa - N +
1)}\frac{\left(\frac{\langle N\rangle}{\kappa }\right)^N}{\left(1
- \frac{\langle N\rangle}{\kappa} \right)^{\kappa - N}},
\label{eq:BD}
\end{equation}
and
\begin{equation}
\frac{1}{\kappa} = 1 - q_N. \label{eq:kappaq}
\end{equation}
For both the NBD and BD we expect the following to hold:
\begin{equation}
\frac{(N+1) P(N+1)}{P(N)} = a + b N \label{eq:dep}
\end{equation}
with
\begin{eqnarray}
a &=& \frac{\langle N\rangle k}{k + \langle N\rangle},\qquad  b
= \frac{a}{k}\qquad {\rm for~NBD}, \label{eq:NBDp}\\
a &=& \langle N\rangle, \qquad \qquad b = 0 ~\qquad {\rm
for~Poisson}, \label{eq:Poisson}\\
a &=& \frac{\langle N\rangle \kappa}{\kappa - \langle N\rangle},
\qquad b = \frac{a}{\kappa}\qquad {\rm for~BD}, \label{eq:BDp}
\end{eqnarray}
\begin{figure}[h]
\includegraphics[width=8cm]{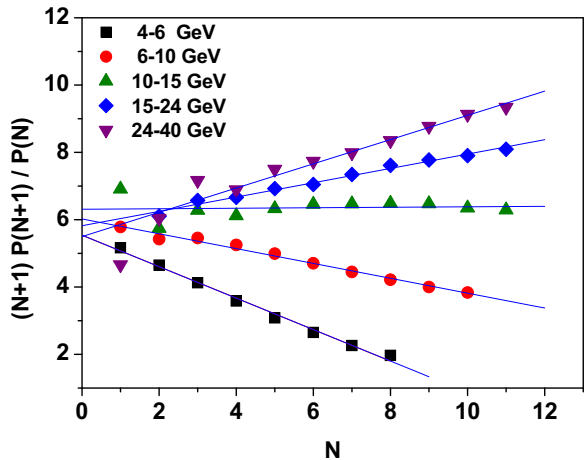}
\vspace{-10mm} \caption{(Color online) $\frac{(N+1)P(N+1}{P(N)}$
in function of multiplicity $N$ in jets with different values of
$p_T^{jet}$ as measured in \cite{ATLAS} and presented in Fig.
\ref{Fig2}. No such information on $P(N)$ is available for jets
analyzed in Fig. \ref{Fig3}. } \label{Fig4}
\end{figure}
\begin{table}[t]
\caption[Table3]{$P(N)$ characteristics for jets with different
$p_T^{jet}$.}
\begin{center}
\begin{tabular}{|r|rrr|}
\hline $p_T^{jet}$ [GeV] &~ $\langle N\rangle $  &~ $Var(N)$~&$q_N - 1$~~ \\
\hline

 4 -  6~~~ &~ 4.41 & 2.31~~  & -0.11~~~ \\
 6 - 10~~~ &~ 5.72 & 3.83~~  & -0.058~~~ \\
10 - 15~~~ &~ 7.11 & 6.61~~  & -0.0098~ \\
15 - 24~~~ &~ 7.56 & 11.2~~  & 0.063~~ \\
24 - 40~~~ &~ 7.80 & 18.1~~  & 0.097~~ \\
\hline
\end{tabular}
\label{Table3}
\end{center}
\end{table}

From data on multiplicity distributions, $P(N)$, measured in jets
\cite{ATLAS} (for $p_T^{jet} \leq 40$ GeV only) one can check the
behavior of  Eq. (\ref{eq:dep}). As can be seen from in Fig.
\ref{Fig4} this relation is linear, i.e., the corresponding $P(N)$
are indeed of NBD or BD type (the deviation from linearity occurs
only for $N=1$, for which one encounters experimental difficulties
and which, in fact, can be omitted from our analysis). From
parameters $a$ and $b$ obtained this way we can deduce, using Eqs.
(\ref{eq:NBDp}) - (\ref{eq:BDp}), values of $\langle N\rangle$,
$Var(N)$ and $k$ or $\kappa$ (i.e., values of the corresponding
nonextensivity parameter $q_N$) which are presented in Table
\ref{Table3}. Notice that their values correspond closely to those
obtained from the distributions of $p_T$ in jets presented in
Table \ref{Table1}.

\section{Self-similarity property of the multiparticle
production processes} \label{sec:Sspmpp}

\begin{figure}[h]
\includegraphics[width=8.5cm]{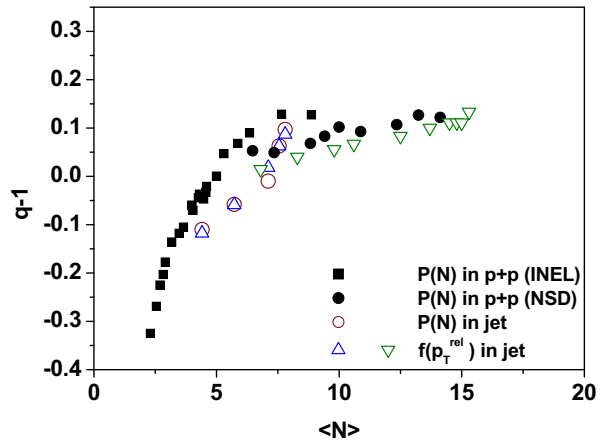}
\vspace{-10mm} \caption{(Color online) Compilation of values of
$q$ as obtained from $p_T^{rel}$ distributions (triangles) and
from multiplicity distributions (circles). Triangles at small
$\langle N\rangle$ are obtained from data \cite{ATLAS}, those for
larger $\langle N\rangle$ from \cite{ATLAS1}. Full squares and
circles are from data on multiparticle production in $p+p$
collisions and, correspondingly, squares (inelastic data) are from
compilation for LAB energy $3.7-303$ GeV presented in \cite{AKW},
whereas circles (non-single diffractive data) are from compilation
presented in \cite{G-G}.} \label{Fig5}
\end{figure}

The values of nonextensivity parameters obtained from an analysis
of multiplicity distributions and distributions of $p_T$ of jets
and in jets can now be compared with the respective nonextensivity
parameters obtained in measurements of $p_T$ distributions in
other experiments on minimum bias $pp$ collisions in which the
range of $p_T$ and multiplicities were similar and energies of
which were similar to energies of the jets investigated. The
corresponding results for the dependence of the resulting
nonextensive parameters $q$ as a function of the measured mean
multiplicity $\langle N\rangle$ are presented in Fig. \ref{Fig5}.
The approximate similarity of these results is clearly
visible\footnote{A word of comment on Fig. \ref{Fig5} is in order
here. So far we were estimating the parameter $q$ from
distributions of $p_T$ or $N$ and discussing its energy
dependence, $q(s)$, as obtained from different experiments
\cite{cond,Addrev1}. Here we would like to compare distributions
of particles in $p+p$ collisions to those in jets, for which,
unfortunately, we do not know the corresponding energy $\sqrt{s}$.
On the other hand, we know $\langle N\rangle$ both for $p+p$
collisions and for particles produced in jets, so it is reasonable
instead to show $q$ as a function of $\langle N\rangle$.}.

The results presented here can be summarized in the following way:
$(i)$ A Tsallis distribution successfully describes inclusive
$p_T$ distributions in a wide range of transverse momenta for all
energies measured so far \cite{cond,WW,CYWGW}. $(ii)$ This is also
true for the distribution of transverse momenta of jets as shown
in Fig. \ref{Fig1}. The nonextensivity parameter in this case, $q
= 1.14$, is comparable to $q = 1.15$ describing inclusive
distributions of transverse momenta of particles at the same
energy $7$ TeV \cite{CYWGW}. $(iii)$ The Tsallis distribution also
describes transverse momenta distributions of particles in jets.
The values of $q$ obtained in this case are roughly the same as
those obtained from an analyzes of multiplicity distributions in
these jets. It should be noted that, as seen in Fig. \ref{Fig5},
values of the nonextensivity parameter $q$ for particles in jets
correspond rather closely to values of $q$ obtained from the
inclusive distributions measured in $pp$ collisions (for the
corresponding energies available for production) in the similar
ranges of transverse momenta\footnote{Results discussed here could
be regarded as related to the phenomenon of geometrical scaling
for $p_T$ distributions discussed recently (cf. \cite{CAn} and
references therein), apparently being a consequence of gluon
saturation at some scale $Q_s$. It turns out that scaled
distributions can be described by a Tsallis formula \cite{RWW}
with the saturation scale being hidden in the parameter $T$ (not
$q$); in fact to get scaling one has to allow for $T$ being
dependent on $p_T$. One should, however, be aware of the fact that
in the energy domain discussed here scaling seems to be violated
\cite{MP}.}.

To conclude, one observes a kind of similarity (in what concerns
values of the corresponding nonextensivity parameters) of
multiplicity distributions $P(N)$ and transverse momentum
distributions $f\left( p_T\right)$ of particles produced in
minimum bias collisions $pp$ and particles in jets of comparable
energies. This can be interpreted as a demonstration that the
mechanisms of particle production in both cases are the same or,
at least, are similar and contain some {\it common part}
\cite{Satz}. This common part, in turn, can be identified with the
self-similarity character of the production process in both cases,
resulting in a kind of cascade process, which always results in a
Tsallis distribution \cite{Kas}. Actually, this is a very old
idea, introduced already by Hagedorn in \cite{SsH}. He assumed
that the production of hadrons proceeds through formation of {\it
fireballs which are a statistical equilibrium of an undetermined
number of all kinds of fireballs, each of which in turn is
considered to be a fireball}. In fact this was used as a
justification in the first proposed generalization of the Hagedorn
model, considered as a statistical model, to $q$-statistics, cf.,
\cite{Beck}. In the pure dynamical QCD approach to hadronization,
one encounters the same idea, as, for example, that presented in
\cite{SsQCD}. In it partons fragment into final state hadrons
through multiple sub-jet production. As a result one has a {\it
self-similar behavior of cascade of jets to sub-jets to
sub-sub-jets \dots to final state hadrons}.

\section{Summary}
\label{sec:SC}

Using the Tsallis nonextensive approach, we have analysed recent
data found by the LHC ATLAS experiment \cite{ATLAS,ATLAS1} on
transverse momentum distributions of jets, particles within jets
and their multiplicity distributions. The values of the respective
nonextensivity parameters obtained this way, when compared with
the corresponding values obtained from the inclusive distributions
measured in $pp$ collisions for the corresponding energies
available for production and in similar ranges of transverse
momenta,  were found to be similar. This can be considered as
strong evidence of the existence of some common mechanism behind
all these processes which we tentatively identify with a {\it
self-similarity} property and cascade type processes based on
multiplicative noise \cite{Kas}. They are known to lead to a
Tsallis distribution (with $n - 2 = \langle \eta
\rangle/Var(\eta)$ given by fluctuations of multiplicative noise
$\eta$ \cite{Kas}) of the same type as those describing
statistical or thermodynamical systems (with $q - 1 =
Var(T)/\langle T\rangle$ given by fluctuations of temperature $T$
\cite{qWW,cond})\footnote{A word of caution should be added. The
observed self-similar behavior of distributions of particles
inside the jets  and particles produced in inelastic pp collisions
can be indications of self-similarity in multiparticle production
processes. Jets being a part of all produced particles are
approximately similar to inelastic collisions. However, in reality
we are not able to observe the whole process of hadronization or
to analyze all its subprocesses to really speak of self-similarity
in multiparticle production processes. We only have information on
one such subprocess, i.e., on the production in jets. We observe
similarities between them and multiparticle production in
innelastic collisions in total. This observation is the basis of
our claim that we are dealing with a process which shows the same
statistical properties at many scales. This is our
self-similarity.}.

It is worth reminding at this point that both Tsallis distribution
and the Negative Binomial Distribution can be regarded as a
consequence of using a gamma distribution for clusters formed
before fragmentation. Whereas the former arises from the
fluctuations of temperature in a Boltzman-Gibs distribution, the
latter arises from the fluctuations of mean multiplicity in a
Poissonian distribution. The common feature is that in both cases
fluctuations are given by a gamma distribution which is stable
under the size distribution, i.e., exhibits self-similarity and
scaling behavior (actually, NBD is also a self-similar
distribution \cite{CaT}). This indicates once more that
self-similarity encountered in processes under consideration is
the physical ground of the observed similarities discusses here.
Results presented here could possibly open discussion about the
validity of thermal models \cite{Bugaev}.

 \vspace*{0.5cm}

\noindent
{\bf Acknowledgments}

\vspace{0.3cm}

This research  was supported in part by the National Science
Center (NCN) under contract DEC- 2011/03/B/ST2/02617 (ZW) and by
the Ministry of Science and Higher Education under contract
DPN/N97/CERN/2009 (GW). We would like to warmly thank Dr Eryk
Infeld for reading this manuscript.

\end{document}